# Deterministic aperiodic composite lattice-structured silicon thin films for photon management

Jolly Xavier*[†], Jürgen Probst*, and Christiane Becker

Helmholtz-Zentrum Berlin für Materialien und Energie GmbH, Kekuléstr. 5, 12489 Berlin, Germany

[†]Present Address: Max Planck Institute for the Science of Light, Guenther-Scharowsky-Str.1, 91058 Erlangen, Germany

*E-mail: jolly.xavier@mpl.mpg.de, juergen.probst@helmholtz-berlin.de

**Exotic manipulation of the flow of photons in nanoengineered semiconductor materials with an aperiodic distribution of nanostructures plays a key role in efficiency-enhanced and industrially viable broadband photonic technologies. Through a generic deterministic nanotechnological route, in addition to periodic, transversely quasicrystallographic or disordered random photonic lattices, here we show scalable nanostructured semiconductor thin films on large area nanoimprinted substrates up to 4cm$^2$ with advanced functional features of aperiodic composite nanophotonic lattices having tailorable supercell tiles. The richer Fourier spectra of the presented artificially nanostructured materials with well-defined lattice point morphologies are designed functionally akin to two-dimensional incommensurate intergrowth aperiodic lattices-comprising periodic photonic crystals and in-plane quasicrystals as subgroups. The composite photonic lattice-structured crystalline silicon thin films with tapered nanoholes or nanocone-nanoholes are presented showing up to +30 % achievable short circuit current density in comparison to a periodic counterpart where as it is up to +190 % in comparison to a reference unstructured silicon thin film of 300 nm thickness. In view of scalable bottom-up integrated device fabrication processes, the structural analysis is further extended to liquid phase crystallized double-side-textured deterministic aperiodic lattice-structured 10 μm thick large area crystalline silicon film.**

The dense Fourier spectra of aperiodic lattice-embedded nanostructured materials strongly modulate and enable flexible tailoring of the light-matter interaction for broad band nanophotonic applications in comparison to unstructured bulk materials [1-10]. The seemingly counter intuitive impact of *order* as well as *disorder* of such nanostructures results in achieving tailorable optical properties of nanostructured materials for integrated photonic applications. Due to their inherent ability to boost broad band absorption independent of angle of incidence and polarization of incident light, deterministic disordered and transversely quasicrystalline-structured semiconductor materials are investigated as appealing option as efficiently absorptive media for integrated nanophotonic device applications [3-5, 10-17]. On the other hand, in order to ensure technological practicability with industrial viability, the predictive models, specific structural engineering design rules as well as high throughput large-area fabrication feasibility of high resolution nanostructured materials are unavoidable and highly demanding [17-20]. Engineering the correlated geometric distribution of the refractive index in nanoscales in artificially structured semiconductor materials would lead to even more viable control on light in-coupling as well as light propagation within the semiconductor materials. Here we report on nanostructured c-Si thin films with effective advanced functional features of tailorable '*deterministic aperiodic composite photonic lattices*', from now on called for brevity as '*composite lattices*'. They are functionally embedded with designed characteristics of multiple lattice periodicities of conventional photonic crystals and in-plane higher order rotational symmetries of common quasicrystals, hence the denomination '*composite lattices*'. By means of advanced nanoengineering, the presented large area composite lattices are deterministically designed and realized functionally similar to two dimensional incommensurate intergrowth aperiodic lattices surpassing the dense distribution of lattice points in real space proportional to the increase in sets of intergrowth elements [6]. We use a generic Fourier reconstruction method involving selective superposition of components to generate easily tailorable irradiance profiles of composite lattices. By a subsequent thresholding and discretization of lattice points within a chosen mesoscopic supercell of typically 10 to 25 µm in lateral dimension in one direction, the possible fabrication limitations of the otherwise dense chosen mesoscopic supercell of typically 10 to 25 µm in lateral dimension in one direction, the possible fabrication limitations of the otherwise dense and overlapping spatial distribution of even superlattice structures in real space are also surpassed. It is also pertinent to be noted that if adopted for substrate structuring using large throughput fabrication processes like nanoimprint lithography, earlier reported deterministic aperiodic lattices with mesoscopic unit cell comprising of nanostructures with structural morphology of very fine and sharp grooves are likely to be drastically modified during overlaying multi-layered photonic device fabrication processes of standard bottom-up approach [17, 18]. This eventually may lead to deviation in their expected scattering as well as light trapping performance in addition to undesired increment in dislocation density in the substrate-silicon interfaces in the final device. So in addition to the designed aperiodic geometric nature of the supercell to facilitate rich Fourier spectra, in the presented nanophotonic composite lattices we have also considered the tailorable number of lattice points and discretized morphology of the basis structure at the individual lattice point which stays intact even during the multilayered deposition. Moreover, as being shown later, the presented approach is also well compatible with state-of-art high quality liquid phase crystallization (LPC) for structured-Si thin films [21]. This opens up new vistas in the field of advanced nanoengineering for broadband integrated applications in solar energy conversion as well as storage, exotic light emitting diodes, slow light integrated chips, spectrally tailorable biosensors, and efficient nanophotonic test beds for nonlinear light-matter interactions like Anderson localization [1, 3-5, 7, 10].

**Design and computational analysis**

Here the nanostructured composite lattice is designed by Fourier reconstruction involving the super position of selective plane waves distributed in $s \geq 2$ sets and subsequent discretization (*see Methods*). Each set has a characteristic length scale and a rotational symmetry $q_m$ (where *m* is from 1 to *s*). In Fig. 1 we show a few examples of composite lattices, classified into two types.

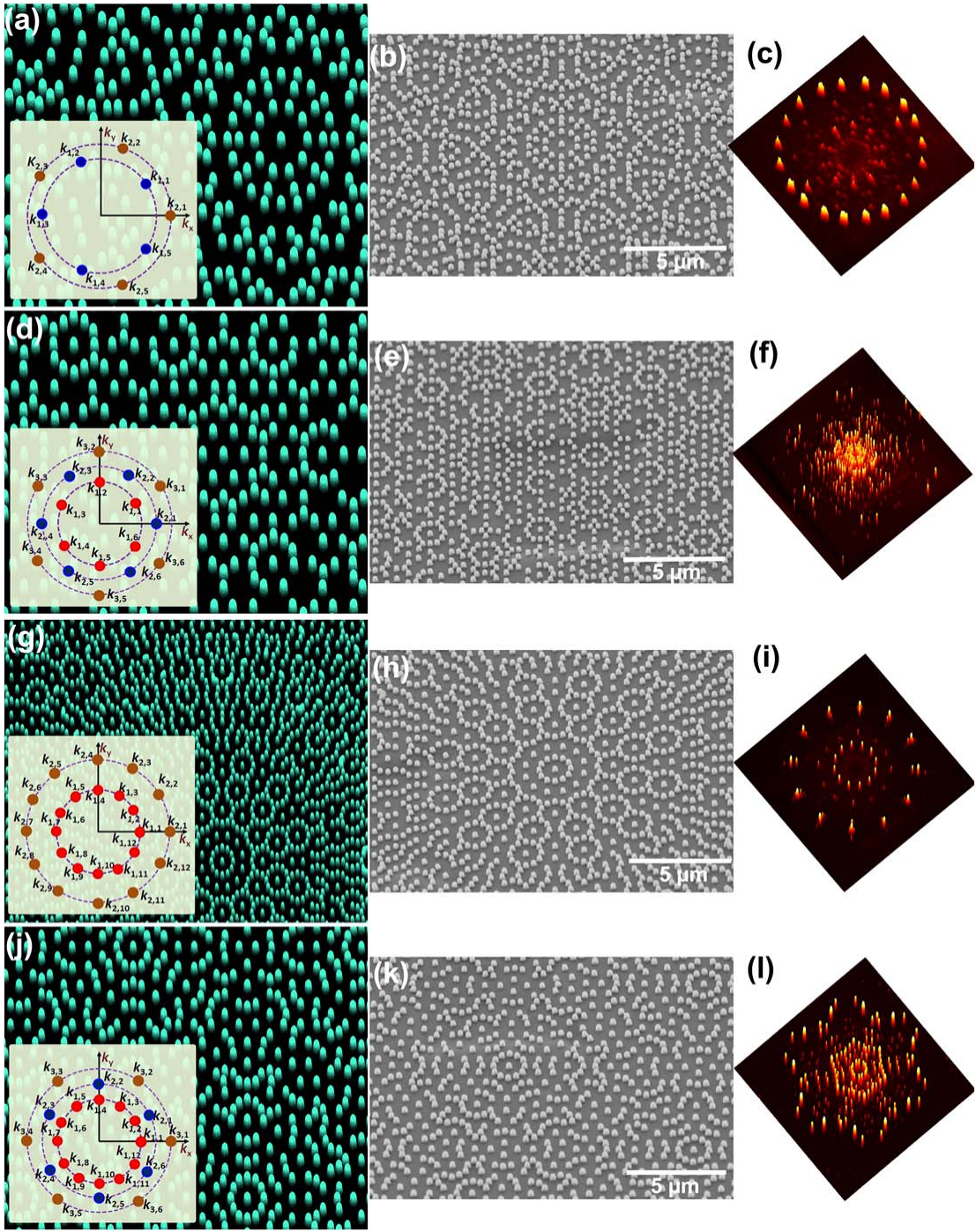

**Figure 1 | Deterministic aperiodic composite nanophotonic lattices.** First column: Part of computed composite lattice structure (Inset: Sets of superposing k-vector component representation in $k_x$-$k_y$ plane), Second column: SEM images (40° tilted) of nanoimprinted substrates, Third Column: Experimentally recorded diffraction pattern intensity distribution. (a)-(c) PPC$_{penta}$ with $s = 2$, $q_1 = q_2 = 5$. (d)-(f) PPC$_{hexa}$ with $s = 3$, $q_1 = q_2 = q_3 = 6$. (g)-(i) PPC$_{dodeca}$ with $s = 2$, $q_1 = q_2 = 12$. (j)-(l) PSC with $s = 3$, $q_1 = 12$, $q_2 = q_3 = 6$.

The first three rows (Figs. 1a-1i) show poly periodic composite (PPC) photonic lattices where the number $q_m$ is same in all sets, starting with a lattice embedded with ten-fold rotational symmetry $q_m$ = 5), a hexagonal PPC (comprised of three sets ($q_m$ = 6) and a PPC with dodecahedral rotational symmetry ($q_m$ = 12). The last row (Figs. 1j-1l) shows a poly symmetry composite (PSC) photonic lattice where the number $q_m$ is not same in all sets, combining both transversely quasicrystallographic 12-fold rotational symmetry structures as well as a mesoscopic hexagonal order. The insets in the first column of Fig. 1 show the sets of *k*-vector components giving the characteristic representation of each design, scanning electron microscopic (SEM) images of experimentally realized nanoimprinted glass substrates and their respective far field diffraction pattern intensity distributions are shown in the second and third columns. In principle any arbitrary number of rotational symmetries can be embedded given the complexity involved in the generated aperiodic pattern. For brevity, conventional periodic and transversely quasicrystallographic photonic lattices can be considered here as subgroups within this broad class of photonic lattices when $s$ = 1 and the respective number $q$ point to their fundamental rotational symmetry.

To picturize the details of our approach we closely consider one of the above cases that of PPC$_{hexa}$ (*see also Methods*), where $s$ = 3 and $q_1$ = $q_2$ = $q_3$ = 6 (inset of Fig. 2a) with absolute amplitude strengths of the components in respective set as 0.5:0.25:1. In the subsequent sections we make comparative computational as well as experimental analysis of this composite photonic lattice with periodic, transversely quasicrystallographic as well as disordered photonic lattices with comparable fill fraction. The considered radial distance from the origin of the components in set 1 corresponds to that of the first order Fourier components of a hexagonal lattice with real space lattice constant $a_1$ = 1000 nm whereas for the components in set 2 and 3, it is respectively $a_2$ = 800 nm and $a_3$ = 600 nm. The mesoscale supercell (shaded region in Fig. 2b) lateral dimensions are respectively $a_{PPC}$ = 10.4 µm and $b_{PPC}=\sqrt{3}a_{PPC}$ = 18.01 µm, such that the complementary edges keep the rotational symmetry of the basic supercell intact even across neighboring tile boundaries. This supercell with 741 lattice points is tiled to get the large area composite lattice.

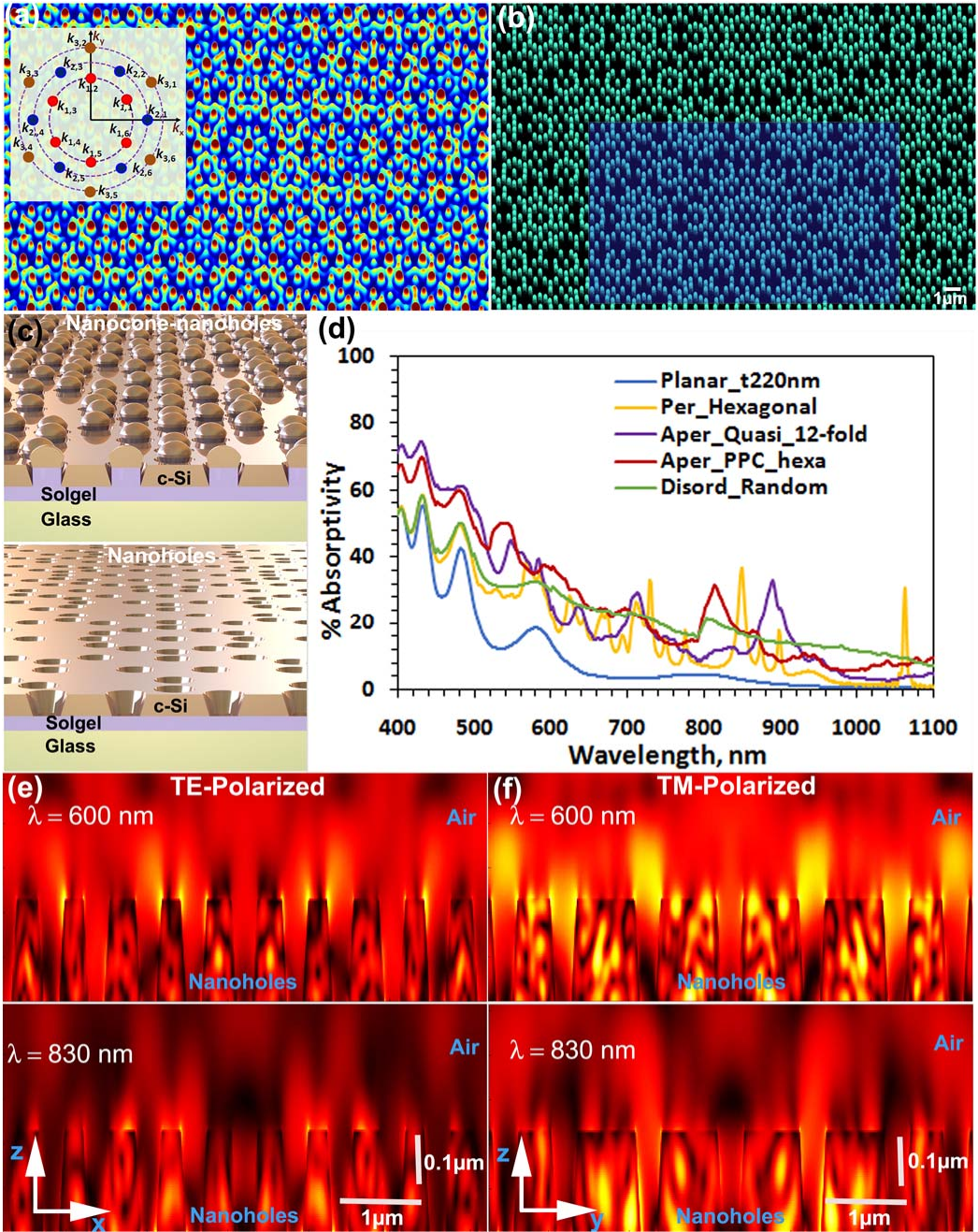

**Figure 2 | Computational analysis of PPC$_{hexa}$ with $s = 3$, $q_1 = q_2 = q_3 = 6$.** (a) Intensity distribution of the basic interference irradiance profile. Inset: Sets of superposing k-vector component representation in $k_x$-$k_y$ plane. (b) Generation of the discretized lattice by tiling the mesoscale supercell shown in blue shadowed region. (c) Schematic of respectively nanocone-nanoholes (top) and nanoholes (bottom) (d) Computed broadband absorption spectra of nanohole structured c-Si films of thickness = 220 nm. (e)-(f) Part of electric field intensity distribution in cross sectional planes along the center of the 10.4 µm x 18.01 µm sized supercell with 741 tapered nanoholes in c-Si thin film.

The individual rod diameter of 236 nm is calculated such that same fill fraction is obtained comparable to periodic, quasicrystallographic and random disordered lattices used in the present study. Given the geometrical and morphological advantages of the tailorable composite lattices, they could very well be designed to simultaneously couple the photon flux in the UV, and trap the visible as well as the infrared region either for broad band applications or for a specific spectral range. We have chosen here Si film of 220 nm thickness structured with tapered nanoholes with a slanting angle of 17° as observed in the experimental scanning images. The tapered nanohole bottom diameter was chosen to be 196 nm where a 15% shrinkage in the solgel nanoimprinted lattice rod diameter is considered as indicated in the experimental analysis. Apart from a planar reference c-Si thin film, for the computational as well as the subsequent experimental comparative analysis in the present study we included c-Si thin films structured with periodic hexagonal lattice, transversely 12-fold symmetry quasicrystal, and disorder-induced random lattice (Fig. S2 *Supplementary material*). Using a 3D Finite Difference Time Domain (FDTD) Maxwell's equation solver (Lumerical) we computed the broadband absorption spectra as given in Fig. 2d. As it is seen, the aperiodic lattices show an enhanced integrated absorption spectrum in the broad band range. The broad band light coupling as well as efficient light trapping of the designed $PPC_{hexa}$ structure is further visualized by computing the cross sectional field intensity distribution of the aperiodic lattice while a plane wave is incident from above. The 3D FDTD computer simulation of the electric field intensity distribution for two wavelength regions at $\lambda$= 600 nm as well as at $\lambda$= 830 nm for TM and TE incident polarizations is given in Figs. 2e and f showing the high field in-coupling and confinement which in turn leads to overall enhanced absorption (Fig. 2d), which will be further proved in the experimental analysis later in this letter. In Figs. 3 and S3 (*see Supplementary material*) we give our computational results for the present aperiodic $PPC_{hexa}$ by tuning the amplitude strength of the three sets of k-vector components whereby the strength of a particular intergrowth pitch could be tailored individually for varying spectral applications without affecting the inherent lattice rotational symmetry or the morphology at the lattice points. Given the thickness of the Si film, as seen in Fig. 3e the absorption spectral response to a range can be tailored without changing the embedded rotational symmetry of the aperiodic lattice. While $PPC_{hexa-2}$ and PPC$_{hexa-3}$ respond well to the smaller wavelengths, PPC$_{hexa-4}$ absorbs comparatively well the longer wavelengths. For the present case, PPC$_{hexa-2}$ has an efficient broadband absorption spectra among all.

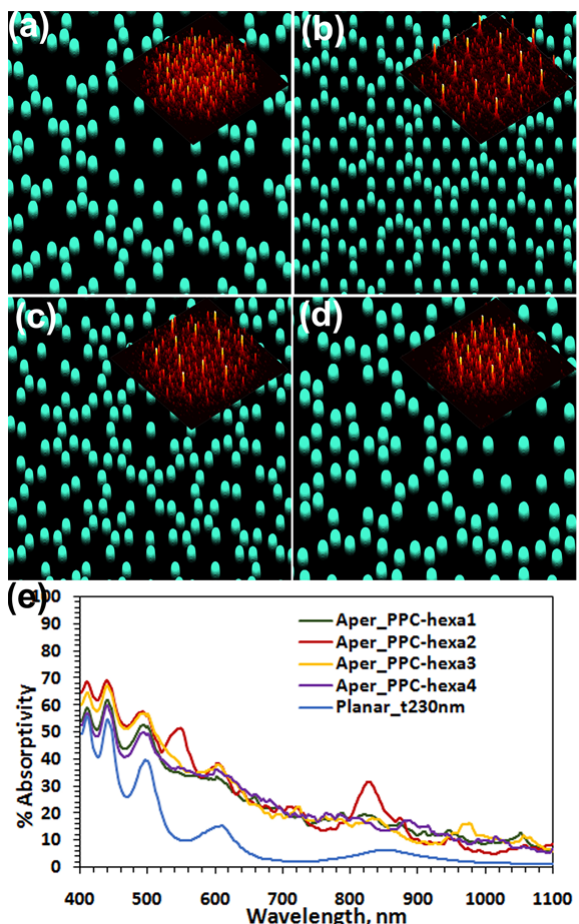

**Figure 3 | Nanoengineering the lattice point distribution.** Tailoring the lattice point distribution of PPC$_{hexa}$ composite lattice with $s = 3$, $q_1 = q_2 = q_3 = 6$ by tuning the ratio of the absolute amplitude strengths of the components in each set. Inset: Resultant Fourier spectrum. (a) PPC$_{hexa-1}$ with ratio 1:1:1. (b) PPC$_{hexa-2}$ with ratio 0.5:0.25:1. (c) PPC$_{hexa-3}$ with ratio 0.25:1:0.5. (d) PPC$_{hexa-4}$ with ratio 1:0.5:0.25. (e) Computed broadband absorption spectra of nanoholes-structured Si thin films (Si thickness = 230 nm).

## Experimental Results and Discussion

Next we analyze experimentally the broad band absorption properties of the structured c-Si thin films on nanoimprinted glass substrates (*see Methods*). Our approach of deterministic fabrication for diverse composite photonic lattices, transversely quasicrystals, periodic lattices and the disordered random lattices are compatible with standard high resolution e-beam lithography for the one-time master wafer fabrication and subsequent large throughput as well as cost effective nanoimprint lithography for highly resolved large area nanostructures [22, 23]. A photographic image of one of the fabricated master wafers is given in inset of Fig. 4a. Nanoimprinted glasses featuring sol-gel rods at each discrete lattice point were used as substrate for the fabrication of silicon nanocone-nanoholes (NCNH) as well as tapered nanohole arrays (NH) (*see Methods*). In Fig. 4a, a SEM image of Si nanodomes (prior to crystallization) with $PPC_{hexa}$ geometry is given. Further we show the subsequently fabricated $PPC_{hexa}$ and 12-fold rotational symmetry NCNH-embedded Si thin films with 300 nm Si deposition thickness (Figs. 4b and 4d respectively) and Si film of 200 nm thickness structured with tapered NH (Figs. 4c and 4e respectively) on nanoimprinted glass substrates. In Fig. S4 (*see Supplementary material*) we give the SEM images of a few fabricated comparable Si thin films structured with NCNH and tapered NH with periodic, transversely quasicrystallographic as well as disordered random lattice geometry realized through the same approach. Figs. 4f and 4g show the respective experimental broadband absorption spectra of them. The 3D visualization schematics of the material layers and their morphologies for NCNH and tapered NH are respectively given in the inset of Figs. 4f and 4g. The experimental absorption spectra again approve the performance of aperiodic nanostructures with tailored PPC lattices and quasicrystallographic lattices for improved light coupling into an absorbing medium which also promises enhanced quasi guided light propagation leading to better light trapping within the semiconductor material due to the tailored rich Fourier spectra of the structurally engineered aperiodic lattices presented here. In order to analyze the sole impact of geometrical and morphological features of the studied nanostructured thin film, we neither used any antireflection coating nor any back reflectors in the present study. The use of such material layers naturally improves further the absorption in the given Si thin films [16, 23]. As an evaluation of the efficient broadband absorption properties of structured Si thin films maximum achievable short circuit current density $j_{sc,max}$ is calculated (*see Methods*).

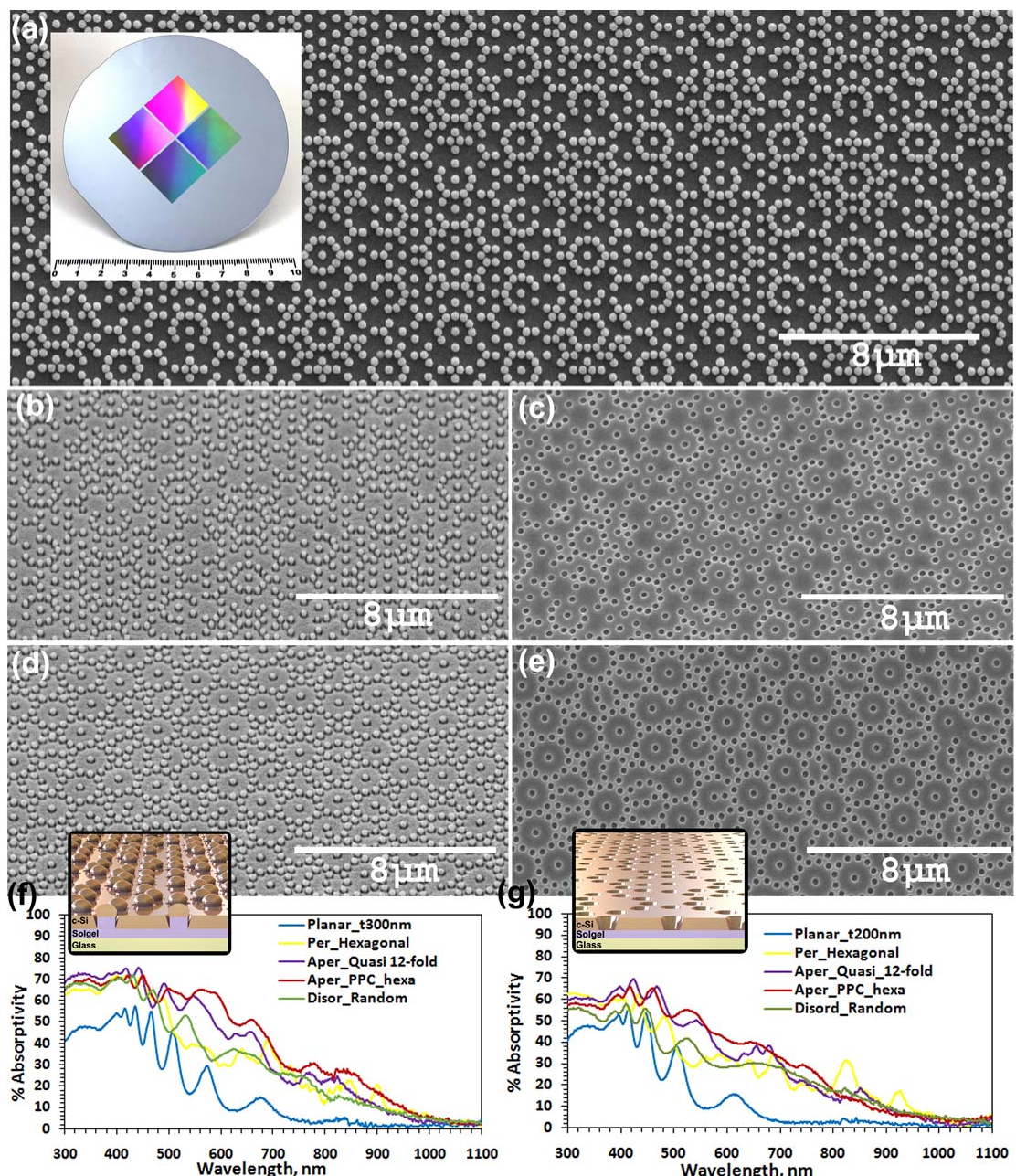

**Figure 4 | Experimental analysis of aperiodic nanostructured c-Si thin films** (a) SEM image of the fabricated large area PPC$_{hexa}$ composite lattice structured thin film with Si nanodomes (prior to crystallization) on nanoimprinted glass substrate. Inset: Photograph of one of the master structures in Si wafer fabricated via e-beam lithography (Scale bar: 10 cm). (b)-(c) SEM images of the PPC$_{hexa}$ composite lattice structured c-Si thin film respectively textured with nanocone-nanoholes (Images are tilted by 40°) (b) and nanoholes (c). (d)-(e) SEM images of the transversely 12-fold symmetry quasicrystal-structured c-Si thin film respectively textured with nanocone-nanoholes (d) and nanoholes (e). (f)-(g) Experimental broadband absorption spectra respectively for the c-Si thin films textured with nanocone-nanoholes (f) and nanoholes (g). Inset: Schematic representation of the respective layered structure.

From Fig. 4f and 4g as well as from Table 1, it can be seen that respectively for the cases of the nanohole and nanocone-nanohole engineered c-Si thin films, the aperiodic lattices especially the presented PPC$_{hexa}$ shows the highest integrated absorption. The enhancement in absorption amounts to a factor of 2.75 and 2.9 respectively for nanoholes and nanocone-nanoholes compared to an unstructured Si thin film. Considering a more meaningful estimation, a +18.36 % and +30 % relative higher maximum achievable short circuit current density $j_{sc,max}$ is estimated for PPC$_{hexa}$ compared to the corresponding value of the periodic hexagonal lattice under investigation and the same trend is observed in comparison to the investigated disordered random lattice too. Whereas these enhancements in comparison to a planar reference were respectively +175.3 % and +189.8 %. For the aperiodic transversely 12-fold symmetry quasicrystal structured samples with NH and NCNH respectively these enhancements in $j_{sc,max}$ were +10.5 % and +14.28 % in comparison to hexagonal lattice. Further analysis of the optical absorption spectra (Figs. 4f and 4g) show that two mechanisms are responsible for increased broadband absorption in the thin c-Si layers: From the short wavelengths regime at λ < 500 nm it can be seen that reflection losses are strongly suppressed in the aperiodic structured silicon films. At longer wavelengths around 650 nm and 900 nm regimes very high absorption occur outperforming the absorption in other lattices. This could be attributed to the large density of Fourier states in the presented aperiodic lattices such that light can effectively be coupled into quasi-guided modes of the thin silicon film. The observed broadband absorption enhancements are as predicted for disordered photonic lattices in the earlier shown numerical studies [16, 17] and verified by our experimental results indicating that aperiodic nanophotonic structures improve even advanced periodic nanophotonic light-trapping concepts [13]. These comparative experimental absorption results strongly indicate that these tailored composite and quasicrystalline aperiodic lattice-embedded artificially nanoengineered materials are excellently suited for broadband absorption enhancement in ultrathin nanophotonic integrated devices. They efficiently exploit in a unified manner the resonant optical properties of periodic lattices as well as the broadband features of the disordered nanophotonic structures. Moreover, through the presented approach their geometric and structural features can be very precisely tailored to suit to the required semiconductor film thickness dependent absorption spectral response as per the application requirement.

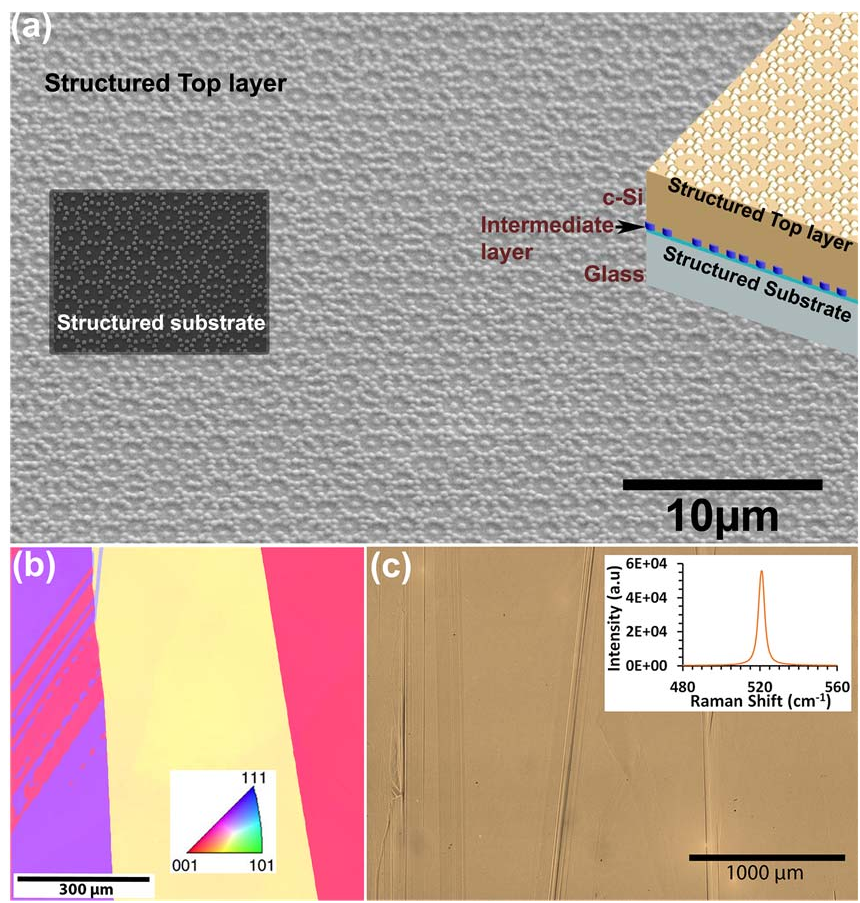

**Figure 5 | Deterministic double side aperiodic lattice-textured c-Si thin films** a) SEM image of the top surface of the layered sample with laser beam-assisted liquid phase crystalized Si film of 10 μm thickness. Left inset: Structured substrate at the bottom. Right inset: Schematic of Glass (bottom most)-Solgel-$SiO_x$-Si-$SiN_x$ layered sample. b) One of the EBSD surface orientation maps of double side aperiodic lattice-structured LPC Si thin film prior to $SiN_x$ deposition. The color coding of the different crystallographic orientations is depicted in the inset. c) Optical microscope large area image of the top surface. Inset: First-order Raman scattering at room temperature from double side aperiodic lattice-structured LPC Si thin film.

Extending further, as a proof of concept we fabricate and analyze a deterministic aperiodic double-side textured 10 μm thick Si films with enhanced material quality via laser beam liquid phase crystallization (LPC) while the geometrical as well as morphological features of the intended aperiodic lattice is well maintained during multiple layer depositions (*see Methods*). In Fig. 5 we show the double side aperiodic lattice textured LPC Si thin film realized through the present approach. Fig. 5a gives the SEM image of the top surface of the Si layer with an effective 80 nm SiN$_x$ layer on top of it. The left inset of Fig 5a gives the SEM image of the bottom structured substrate on which the subsequent multi layers are deposited, while a schematic of the layered structure is given in the right inset. Further material structural quality analysis are done using electron backscatter diffraction (EBSD) microstructural-crystallographic orientation mapping (Fig. 5b), optical microscope large area imaging and Raman scattering spectral imaging (Fig. 5c). Fig. 5b gives one of the EBSD surface orientation mappings, exhibiting elongated large grains of millimeters range, of a LPC c-Si thin film sample with a 100 nm SiO$_x$ intermediate layer on nanoimprinted glass substrate. The structural characterizations verify the material structural quality of the deterministic aperiodic structured 10 µm thick LPC Si films with well-maintained morphological features. If needed the lattice point morphology of the substrate structuring can be even further shaped with ease from columnar to smoothed hills in order to tune the electrical material quality of c-Si thin film for varied applications without compromising the designed aperiodic geometrical features. Moreover, our generic approach for nanoengineered deterministic aperiodic lattices with rich Fourier spectra offers a great flexibility to tailor their geometry, lattice point morphology, and spectral response range to match to any desired different thin film thicknesses, material of interest and for intended nanophotonic applications. The industrial application-viable and standard bottom-up fabrication technologies-compatible approach which we have demonstrated is envisaged to suit well for broadband integrated nanophotonic devices which demands large throughput and high resolution structured semiconductor materials with advanced tailorable features in nanoscale [1-5].

## Methods

### i) Design approach

Irradiance profile for various transversely quasicrystalline photonic lattice structures could be computed by Fourier reconstruction resulting from the designed superposition of plane waves [9]. For the subwavelength nanophotonic structures, the discreteness of the neighboring lattice points are strongly affected as the number of *k*-vector components are increased [18]. It will be very critical when the components are distributed in multiple sets with varying radial distance from the origin in the $k_x$-$k_y$ Fourier plane. A generalized expression used here to compute the irradiance profile for a desired subwavelength composite photonic lattice structure by Fourier reconstruction resulting from the superposition of $p = \sum_{m=1}^{s} q_m$ plane waves linearly polarized in same direction and distributed in *s* sets is given by,

$$\mathbf{I}(\mathbf{r}) = \left\{ \sum_{i=1}^{p} |\mathbf{E}_i|^2 + \sum_{i=1}^{p} \sum_{\substack{j=1 \\ j \neq i}}^{p} (\mathbf{E}_i \cdot \mathbf{E}_j^*) exp[\mathrm{i}(\mathbf{k}_i - \mathbf{k}_j) \cdot \mathbf{r} + \mathrm{i}(\varphi_i - \varphi_j)] \right\} \quad (1)$$

where, $\mathbf{E}_i$, $\mathbf{k}_i$, $\mathbf{r}$ and $\varphi_i$ are respectively the complex amplitudes, the wave vectors, the position vector and the initial phase which is considered to be zero here. Each set is comprising of $q_m$ (where *m* is from 1 to *s*) azimuthally equidistant components in the $k_x$-$k_y$ Fourier plane. The components in a given set are radially equidistant from the origin designed as per the desired neighbor distance of a constituent lattice structure to be embedded within the composite lattice. The $k_{m,n}$ component in k-vector diagrams (inset of Fig 1) represents the $n^{th}$ component in $m^{th}$ set. The value of $q_m$ in each set basically defines its contributing fundamental rotational symmetry within the composite lattice. With reference to the k-vector representation shown in the inset of Fig. 2a, for the case of PPC$_{hexa}$ we used $p = 18$ where $s = 3$ and $q_1 = q_2 = q_3 = 6$. The radial distance from the origin of the components in the first set corresponds to that of the first order Fourier components of hexagonal lattice with real space lattice constant $a_1 = 1000$ nm whereas the corresponding values for components in second and third sets respectively are $a_2 = 800$ nm and $a_3 = 600$ nm. This gives a means of combining the functional features of three effective pitches within a composite photonic lattice, but at the mean time the real space spatial constraints of accommodating multiple periodicities is alleviated. Further, the strength of amplitude of waves in each of these three sets can be tailored independently and in the present case we have chosen a ratio 0.5:0.25:1. Tuning this ratio is an additional degree of freedom to tailor the composite PPC (Fig. 3). The computed aperiodic lattice pattern is quite large in spatial extension which makes it inappropriate to realize such complex lattices by standard high resolution nanofabrication methods. So in a subsequent step, a large rectangular supercell is chosen from the computed pattern. Next, the pattern within the supercell is discretized by applying a certain threshold to the continuous function, determining the geometric centroids of the area above the set threshold and subsequently tiling the supercell to generate a large area composite lattice (Fig. 2b). Here, we chose a threshold intensity as well as bright spot perimeter equal to 0.3 times of peak values within the supercell. By tuning these threshold values, the number of discrete lattice points within the supercell can be easily tuned without affecting the rotational symmetry properties of the generated lattice to a great extent. The discretized supercell should contain sufficient number of lattice points permitting a discrete Fourier pattern with high finesse. At the mean time the dimensions of the supercell are chosen to be compatible enough both for 3D computational analysis as well as standard high resolution fabrication approaches such as electron-beam lithography. The lateral dimensions of the supercell (blue shaded region in Fig. 2b) of this designed PPC$_{hexa}$ are respectively $a_{PPC}$ = 10.4 µm and $b_{PPC}=\sqrt{3}a_{PPC}$ = 18.01 µm and have complementary edges keeping the rotational symmetry of the basic supercell intact over the whole tiled region. This supercell has 741 lattice points and is subsequently tiled to realize the large area composite photonic lattice. Fig. S1 gives an example on choice of supercell size and its effect on the resultant Fourier pattern, where a 10-fold transversely quasicrystal lattice supercell is optimized with lateral dimensions of 9.4 × 11.05 µm$^2$ (*see Supplementary material*). In Fig. S2 (*see Supplementary material*) we outline the wave design k-vector diagrams, the nanoimprinted substrates and the respective far-field diffraction patterns of large area periodic square and hexagonal lattices as well as transversely quasicrystals with 10-fold and 12-fold rotational symmetries. For a fair comparison it is needed to realize non-overlapping nanoimprinted lattice structures for all the considered photonic lattices with equal fill fractions. So we have chosen a corresponding lattice constant of 700 nm of the simplest periodic structure with square lattice as the reference lattice constant considering the fact that comparable aperiodic lattices would result in an even smaller nearest neighbor distance approaching 400 nm. All the lattices given in Fig. S2 are designed with equal fill fraction as well as comparable lattice constant or nearest neighbor distance indicated by their equal radial distance of the prominent diffraction order from the origin. We also designed a supercell (~14.5 µm x 14.0 µm) of a randomly disordered lattice which was fundamentally generated by transversely perturbing the in-plane Cartesian coordinates of the lattice points of the corresponding periodic hexagonal lattice with $a_{hex}$ = 808 nm used in our study by an amount chosen randomly from a uniform distribution of values between 0 and 0.3 $a_{hex}$.

*ii) Nanostructured master wafer by electron-beam lithography*

Structured silicon master wafers were prepared by electron beam lithography, mask inversion by a lift-off process and final reactive ion etching. Each of the presented photonic lattice pattern was exposed by a 100 kV Vistec EBPG5000+ES onto an area of 2.0 cm x 2.1 cm on a silicon substrate coated with a 160 nm thick layer of positive electron beam resist ZEP520A (ZEON Corp.). After development in hexyl acetate and rinsing in isopropyl alcohol, the structured resist was coated with a 30 nm thick nickel layer, which served as etch mask after resist removal in a lift-off procedure in dimethyl formamide. Finally, the silicon was etched 370 nm deep in a highly anisotropic reactive ion etching process (etching gases $SF_6$, $C_4F_8$ and $O_2$) and the nickel mask was subsequently removed in hydrochloric acid.

*iii) Nanoimprinted Glass substrates*

After a short oxygen plasma pre-treatment, the master wafer with nanostructures is coated with a thin antisticking layer via vapor deposition of (tridecafluoro-1,1,2,2-tetrahydrooctyl) trichlorosilane short F13-TCS (CAS: 78560-45-9, ABCR GmbH) in a vacuum oven. As per the requirement, for high resolution nanoimprint process we prepared composite poly-(dimethyl) siloxane (PDMS) stamps or else conventional soft PDMS stamps were used. Composite PDMS (c-PDMS) stamps are composed of two layers: a thin hard PDMS (h-PDMS) layer supported by a thick soft PDMS (s-PDMS) layer. For the c-PDMS stamp preparation [24], first the h-PDMS is made by mixing step by step 3.4g of a vinylsiloxane prepolymer (CAS 67762-94-1, ABCR GmbH) with 0.016g of platinum-divinyltetramethyldisiloxane as catalyst (CAS 68478-92-2, ABCR GmbH), 0.1g of 2,4,6,8-tetramethyl-2,4,6,8-tetravinylcyclotetrasiloxane (CAS 68478-92-2, Sigma-Aldrich Corp.) as modulator and 1g of a hydrosiloxane prepolymer (CAS 68037-59-2, ABCR GmbH). To this mixture 2 g of Toluene was blended as solvent. This final h-PDMS mixture was spin coated directly on the master wafer and left for 1 hr and subsequently cured for 1 hr at 60°C. Now, the s-PDMS was prepared by mixing the poly-(dimethyl) siloxane and its catalyst (PDMS, Elastosil RT A/B 601 from Wacker) with a ratio of 9:1. On the Master wafer with the thin of h-PDMS layer on top, the degassed s-PDMS is poured and subsequently slowly cured in an oven at 35°C for overnight. The final c-PDMS stamp can then be gently peeled off from the master and used as a mold for the replication process. The sol-gel (Philips) is now spin coated on cleaned glass substrates (Corning Eagle). The PDMS stamp is placed on the sol-gel coated glass substrate and cured with a UV ($\lambda$ = 400 nm) lamp for about 6 minutes. After UV-curing, the PDMS stamp is carefully peeled off the substrate leaving the imprinted nanostructures corresponding to the master wafer. After a quick post-bake for 8 minutes at 100°C, the nanoimprinted glass substrates are further thermally annealed at 600°C for an hour in order to ensure the thermal stability of the sol-gel textured glass substrate by effusing out the remaining organic residues.

iv) *Fabrication of structured silicon thin films*

For the fabrication of nanocone-nanoholes (NCNH) and nanoholes (NH) structured Si thin films, a Si layer of the desired thickness is deposited by electron-beam evaporation in amorphous phase on the cleaned nanoimprinted glass substrates at a substrate-temperature of 300°C. Subsequent thermal annealing at 600°C for several hours in a tube furnace within an $N_2$ atmosphere results in solid phase crystallization of the silicon film aside from the columnar substrate features. The still amorphous Si parts around the columns are removed by wet-chemical etching in an etch solution consisting of concentrated $HNO_3$ (65%, 30 parts), concentrated $H_3PO_4$ (85%, 10 parts), HF (50%, 1 part) and $H_2O$ (43 parts) resulting in nanocone-nanoholes [25]. The nanoholes are realized by mechanically detaching the sol-gel columns by abrasion.

For the fabrication of the presented double-side textured liquid phase crystallized Si thin films, first a SiO$_x$ diffusion barrier layer of around 100 nm thickness is deposited on the cleaned nanoimprinted glass substrates by physical vapour deposition using a reactive magnetron sputtering system. Subsequently the desired thickness of nominally intrinsic nanocrystalline Si was deposited by e-beam evaporation at a substrate-temperature of 600°C. Subsequently, a 250 nm thick SiO$_x$ capping layer was deposited on top of the layer stack by reactive magnetron sputtering. This capping layer prevents the surface of the Si melt from levelling out during crystallization and thus allows maintaining the initial topography of the deposited layer akin to that of the nanoimprinted substrate texturing [26]. Moreover, this cap layer prevents from the dewetting of the liquid Si thin film as well. Then the samples were crystallized using a line shaped cw laser (λ = 808 nm ± 10 nm) from LIMO GmbH [27]. The used optical intensity for the crystallization process was around 3.58 kW/cm². Crystallization was performed at a substrate temperature of 600°C and by moving the samples under the laser beam with a velocity of 11 mm/s. Further, the SiO$_x$ capping layer at the surface was removed using a 5% HF solution as etchant in a wet etching process resulting in a double side textured Si thin film with well-maintained morphology of the nanoimprinted substrate.

*v)* ***Absorption measurement and short circuit current density calculation***

The absorption spectra of the nano-patterned crystalline silicon thin films were measured inside an integrating sphere of a PerkinElmer1050 optical spectrometer with wavelength resolution of 2 nm. Samples were tilted by 8 degrees to ensure that the specularly reflected light does not escape through the entrance port. The maximum achievable short circuit current density $j_{sc,max}$ is calculated by,

$$j_{sc,max} = e \int_{300nm}^{1100nm} A(\lambda) \Phi_{AM1.5g}(\lambda) \, d\lambda \qquad (2)$$

with elementary charge *e*, absorption *A*(λ) and Φ$_{AM1.5g}$(λ) the photon flux of the AM1.5g solar irradiance spectrum. It represents the upper limit of the current density of photovoltaic devices at zero voltage at illumination with the global solar irradiance spectrum AM1.5g and serves as figure of merit for integrated absorption. Here, it is assumed that each photon absorbed in the spectral range from 300 to 1100 nm in the nanostructured silicon film contributes to the solar cell current and no electrical losses occur, e.g. by recombination.

**References**

[1] Priolo, F., Gregorkiewicz, T., Galli, M., & Krauss, T. F. Silicon nanostructures for photonics and photovoltaics. *Nature Nanotechnology* **9,** 19-32 (2014).

[2] Vynck, K., Burresi, M., Riboli, F. & Wiersma, D. S. Photon management in two-dimensional disordered media. *Nature Materials* **11,** 1017-1022 (2012).

[3] Wiersma, D. S. Disordered photonics. *Nature Photonics* **7,** 188-196 (2013).

[4] Dal Negro, L. & Boriskina, S. Deterministic aperiodic nanostructures for photonics and plasmonics applications. *Laser & Photonics Reviews* **6,** 178-218 (2012).

[5] Vardeny, Z. V., Nahata, A. & Agrawal, A. Optics of photonic quasicrystals. *Nature Photonics* **7,** 177-187 (2013).

[6] Smaalen, S. V. Incommensurate crystal structures, *Cryst. Rev.* **4,** 79-202 (1995).

[7] Levi, L. et al. Disorder-enhanced transport in photonic quasicrystals. *Science* **332,** 1541-1544 (2011).

[8] Zoorob, M. E., Charlton, M. D. B., Parker, G. J., Baumberg, J. J., & Netti, M. C. Complete photonic bandgaps in 12-fold symmetric quasicrystals. *Nature* **404,** 740-743 (2000).

[9] Xavier, J., Boguslawski, M., Rose, P., Joseph, J. & Denz, C. Reconfigurable optically induced quasicrystallographic three-dimensional complex nonlinear photonic lattice structures. *Adv. Mater.* **22,** 356-360 (2010).

[10] Liu, J. et al. Random nanolasing in the Anderson localized regime. *Nature Nanotechnology* **9,** 285-289 (2014).

[11] Oskooi, A. et al. Partially disordered photonic-crystal thin films for enhanced and robust photovoltaics. *Appl. Phys. Lett.* **100,** 181110 (2012).

[12] Pratesi, F., Burresi, M., Riboli, F., Vynck, K. & Wiersma, D. S. Disordered photonic structures for light harvesting in solar cells. Opt. Express **21,** A460-A468 (2013).

[13] Paetzold, U. W. et al. Disorder improves nanophotonic light trapping in thin-film solar cells. Applied Physics Letters **104,** 131102 (2014).

[14] Xavier. J. et al. Quasicrystalline-structured light harvesting nanophotonic silicon films on nanoimprinted glass for ultra-thin photovoltaics. *Opt. Mater. Express* **4,** 2290-2299 (2014).

[15] Ren, H., Du, Q. G., Ren, F., & PNG, C. E., Photonic quasicrystal nanopatterned silicon thin film for photovoltaic applications. *J. Opt.* **17,** 035901 (2015).

[16] Bozzola, A., Liscidini, M. & Andreani, L. C. Broadband light trapping with disordered photonic structures in thin-film silicon solar cells. *Prog. Photovolt: Res Appl.* **22,** 1237-1245 (2014).

[17] Martins, E. R. et al. Deterministic quasi-random nanostructures for photon control. *Nature Comm.* **4,** 2665 (2013).

[18] Lubin, S. M., Zhou, W., Hryn, A. J., Huntington, M. D. & Odom, T. W. High-rotational symmetry lattices fabricated by moiré nanolithography. *Nano Lett.* **12,** 4948-4952 (2012).

[19] Lee, T., Parker, G., Zoorob, M., Cox, S. & Charlton, M. Design and simulation of highly symmetric photonic quasi-crystals. *Nanotechnology* **16,** 2703-2706 (2005).

[20] Petti, L. et al. High Resolution Lithography as a Tool to Fabricate Quasiperiodic Crystals. *AIP Conference* Proceedings **1176,** 146-148.

[21] Becker, C. et al. Polycrystalline silicon thin-film solar cells: Status and perspectives. *Solar Energy Materials & Solar Cells* **119,** 112-123 (2013).

[22] Guo, L. J. Nanoimprint lithography: methods and material requirements. *Adv. Mater.***19,** 495-513 (2007).

[23] Becker, C. et al. 5 × 5 cm$^2$ silicon photonic crystal slabs on glass and plastic foil exhibiting broadband absorption and high-intensity near-fields. *Scientific Reports* **4,** 5886 (2014).

[24] Kang, H., Lee, J., Park, J., & Lee, H. H. An improved method of preparing composite poly(dimethylsiloxane) moulds. *Nanotechnology* **17,** 197-200 (2006).

[25] Sontheimer, T. et al. Large-area fabrication of equidistant free-standing Si crystals on nanoimprinted glass. *Phys. Status Solidi RRL* **5,** 376-378 (2011).

[26] Becker, C., Preidel, V., Amkreutz, D., Haschke, J., & Rech. B., Double-side textured liquid phase crystallized silicon thin-film solar cells on imprinted glass. *Solar Energy Materials & Solar Cells*. **135,** 2-7 (2015).

[27] Kühnapfel, S., Nickel, N. H., Gall, S., Klaus, M., Genzel, M. C., Rech, B., & Amkreutz, D. Preferential {100} grain orientation in 10 micrometer-thick laser crystallized multicrystalline silicon on glass. *Thin Solid Films* **576,** 68-74 (2015).

**Acknowledgements**

This research of the Nano-SIPPE group at Helmholtz-Zentrum Berlin is supported by The German Federal Ministry of Education and Research (BMBF) in the program NanoMatFutur (No. 03X5520). The authors thank C. Klimm for SEM as well as EBSD imaging of the fabricated nanostructures. The authors also gratefully acknowledge the support in fabrication, analysis and visualization from C. Hülsen, S. Kühnapfel, C. Barth, M. Krüger, H. Rhein, M. Muske, I. Rudolph and V. Preidel of Helmholtz-Zentrum Berlin.

**Author Contributions**

J.X. conceived the idea and computed the master nanostructures for fabrication. J.P. fabricated the master nanostructures via electron-beam lithography. Subsequently J.X. fabricated the nanoimprinted samples as well as structured silicon thin films and carried out computational analysis as well as optical measurements. C.B. supervised the project. J.X and C.B prepared the manuscript. All authors discussed on the results and reviewed the manuscript.

**Competing Financial Interests statement**

The authors declare no competing financial interests.

**Figure Legends**

**Figure 1** | **Deterministic aperiodic composite nanophotonic lattices.** First column: Part of computed composite lattice structure (Inset: Sets of superposing k-vector component

representation in $k_x$-$k_y$ plane), Second column: SEM images (40° tilted) of nanoimprinted substrates, Third Column: Experimentally recorded diffraction pattern intensity distribution. (a)-(c) $PPC_{penta}$ with $s = 2$, $q_1 = q_2 = 5$. (d)-(f) $PPC_{hexa}$ with $s = 3$, $q_1 = q_2 = q_3 = 6$. (g)-(i) $PPC_{dodeca}$ with $s = 2$, $q_1 = q_2 = 12$. (j)-(l) PSC with $s = 3$, $q_1 = 12$, $q_2 = q_3 = 6$.

**Figure 2 | Computational analysis of $PPC_{hexa}$ with $s = 3$, $q_1 = q_2 = q_3 = 6$.** (a) Intensity distribution of the basic interference irradiance profile. Inset: Sets of superposing k-vector component representation in $k_x$-$k_y$ plane. (b) Generation of the discretized lattice by tiling the mesoscale supercell shown in blue shadowed region. (c) Schematic of respectively nanocone-nanoholes (top) and nanoholes (bottom) (d) Computed broadband absorption spectra of nanohole structured c-Si films of thickness = 220 nm. (e)-(f) Part of electric field intensity distribution in cross sectional planes along the center of the 10.4 µm x 18.01 µm sized supercell with 741 tapered nanoholes in c-Si thin film.

**Figure 3 | Nanoengineering the lattice point distribution.** Tailoring the lattice point distribution of $PPC_{hexa}$ composite lattice with $s = 3$, $q_1 = q_2 = q_3 = 6$ by tuning the ratio of the absolute amplitude strengths of the components in each set. Inset: Resultant Fourier spectrum. (a) $PPC_{hexa-1}$ with ratio 1:1:1. (b) $PPC_{hexa-2}$ with ratio 0.5:0.25:1. (c) $PPC_{hexa-3}$ with ratio 0.25:1:0.5. (d) $PPC_{hexa-4}$ with ratio 1:0.5:0.25. (e) Computed broadband absorption spectra of nanoholes-structured Si thin films (Si thickness = 230 nm).

**Figure 4 | Experimental analysis of aperiodic nanostructured c-Si thin films** (a) SEM image of the fabricated large area $PPC_{hexa}$ composite lattice structured thin film with Si nanodomes (prior to crystallization) on nanoimprinted glass substrate. Inset: Photograph of one of the master structures in Si wafer fabricated via e-beam lithography (Scale bar: 10 cm). (b)-(c) SEM images of the $PPC_{hexa}$ composite lattice structured c-Si thin film respectively textured with nanocone-nanoholes (Images are tilted by 40°) (b) and nanoholes (c). (d)-(e) SEM images of the transversely 12-fold symmetry quasicrystal-structured c-Si thin film respectively textured with nanocone-nanoholes (d) and nanoholes (e). (f)-(g) Experimental broadband absorption spectra respectively for the c-Si thin films textured with nanocone-nanoholes (f) and nanoholes (g). Inset: Schematic representation of the respective layered structure.

**Figure 5 | Deterministic double side aperiodic lattice-textured c-Si thin films** a) SEM image of the top surface of the layered sample with laser beam-assisted liquid phase crystallized Si film of 10 µm thickness. Left inset: Structured substrate at the bottom. Right inset: Schematic of Glass (bottom most)-Solgel-$SiO_x$-Si-$SiN_x$ layered sample. b) One of the EBSD surface orientation maps of double side aperiodic lattice-structured LPC Si thin film prior to $SiN_x$ deposition. The color coding of the different crystallographic orientations is depicted in

the inset. c) Optical microscope large area image of the top surface. Inset: First-order Raman scattering at room temperature from double side aperiodic lattice-structured LPC Si thin film.

**Table 1 | Summary of broadband absorption characteristics of investigated structured c-Si thin films**

| Nanoholes (Silicon reference thickness = 200 nm) | | | | | |
|---|---|---|---|---|---|
| **Lattice structure** | **Planar** | **Hexagonal** | **12-fold quasi** | **PPC$_{hexa}$** | **Disordered random** |
| **J$_{sc\ max}$ (mA/cm$^2$)** | 4.7 | 10.9 | 12.1 | 12.9 | 10.4 |
| **Enhancement (versus hexagonal)** | -57.0 % | - | **+10.2 %** | **+18.4 %** | -4.5 % |
| **Enhancement (versus planar)** | - | +132.6 % | **+158.9 %** | **+175.3 %** | +122.2 % |
| **Nanocone-nanoholes (Silicon reference thickness = 300 nm)** | | | | | |
| **Lattice structure** | **Planar** | **Hexagonal** | **12-fold quasi** | **PPC$_{hexa}$** | **Disordered random** |
| **J$_{sc\ max}$ (mA/cm$^2$)** | 5.6 | 12.5 | 14.3 | 16.3 | 12.3 |
| **Enhancement (versus hexagonal)** | -55.1 % | - | **+14.3 %** | **+30.0 %** | -2.1 % |
| **Enhancement (versus planar)** | - | +122.9 % | **+154.7 %** | **+189.8 %** | +118.2 % |

# Deterministic aperiodic composite lattice-structured silicon thin films for photon management

Jolly Xavier*[†], Jürgen Probst*, and Christiane Becker

Helmholtz-Zentrum Berlin für Materialien und Energie GmbH, Kekuléstr. 5, 12489 Berlin, Germany

[†]Present Address: Max Planck Institute for the Science of Light, Guenther-Scharowsky-Str.1, 91058 Erlangen, Germany

**Supplementary Material**

**Figure S1** | Lattice point distribution within the chosen supercell of studied 10-fold symmetry transversely quasicrystalline lattice and its effect on the finesse of the resultant Fourier spectral peaks. Inset on top left: *k*-vector component representation in $k_x$-$k_y$ plane. (a)-(f) Discretized super cell (top) and its respective Fourier spectra (bottom). Inset: Original irradiance profile. (a) Supercell with 30 lattice points. (b) Super cell with 40 lattice points. (c) Super cell with 62 lattice points. (d) Super cell with 101 lattice points. (e) Super cell with 107 lattice points. (f) Super cell with 271 lattice points (used for subsequent fabrication).

**Figure S2** | Fabricated large-area periodic, and transversely quasicrystallographic photonic lattices. The photonic lattices have been designed such that first prominent Fourier order is diffracted at the same angle from the origin (blue scale bar). The diameters of the nanopillars have been calculated for a resultant equal fill fractions for all the studied lattice structures. First column: *k*-vector component representation in $k_x$-$k_y$ plane. Second column: SEM images of photonic lattice nanoimprinted substrates with equal fill fraction Third Column: Experimentally recorded diffraction pattern while a 532 nm laser beam is incident on the nanoimprinted sample. (a)-(c) Periodic square photonic lattice. (d)-(f) Periodic hexagonal photonic lattice. (g)-(i) Transversely 10-fold symmetry quasicrystalline photonic lattice. (j)-(l) Transversely 12-fold symmetry quasicrystalline photonic lattice.

**Figure S3** | Computed lattice point distribution of PPC$_{hexa}$ composite lattice with $s = 3$, $q_1 = q_2 = q_3 = 6$ while tuning the ratio of the absolute amplitude strengths of the components in each set. Inset: Resultant Fourier spectrum (a) PPC$_{hexa-2}$ with ratio 0.5:0.25:1. (b) PPC$_{hexa-3}$ with ratio 0.25:1:0.5. (c) PPC$_{hexa-4}$ with ratio 1:0.5:0.25. (d) PPC$_{hexa-5}$ with ratio 0.25:0.5:1. (e) PPC$_{hexa-6}$ with ratio 1:0.25:0.5. (f) PPC$_{hexa-7}$ with ratio 0.5:1:0.25.

**Figure S4** | SEM images of the fabricated large area c-Si thin film structured respectively with nanocone-nanoholes (Left column) (40° tilted images) and nanoholes (Right column), with lattice symmetry of (a)-(b) Periodic hexagonal lattice, (c)-(d) Transversely 10-fold symmetry quasicrystalline lattice, and (e)-(f) Disordered random lattice.

**Figure S1**

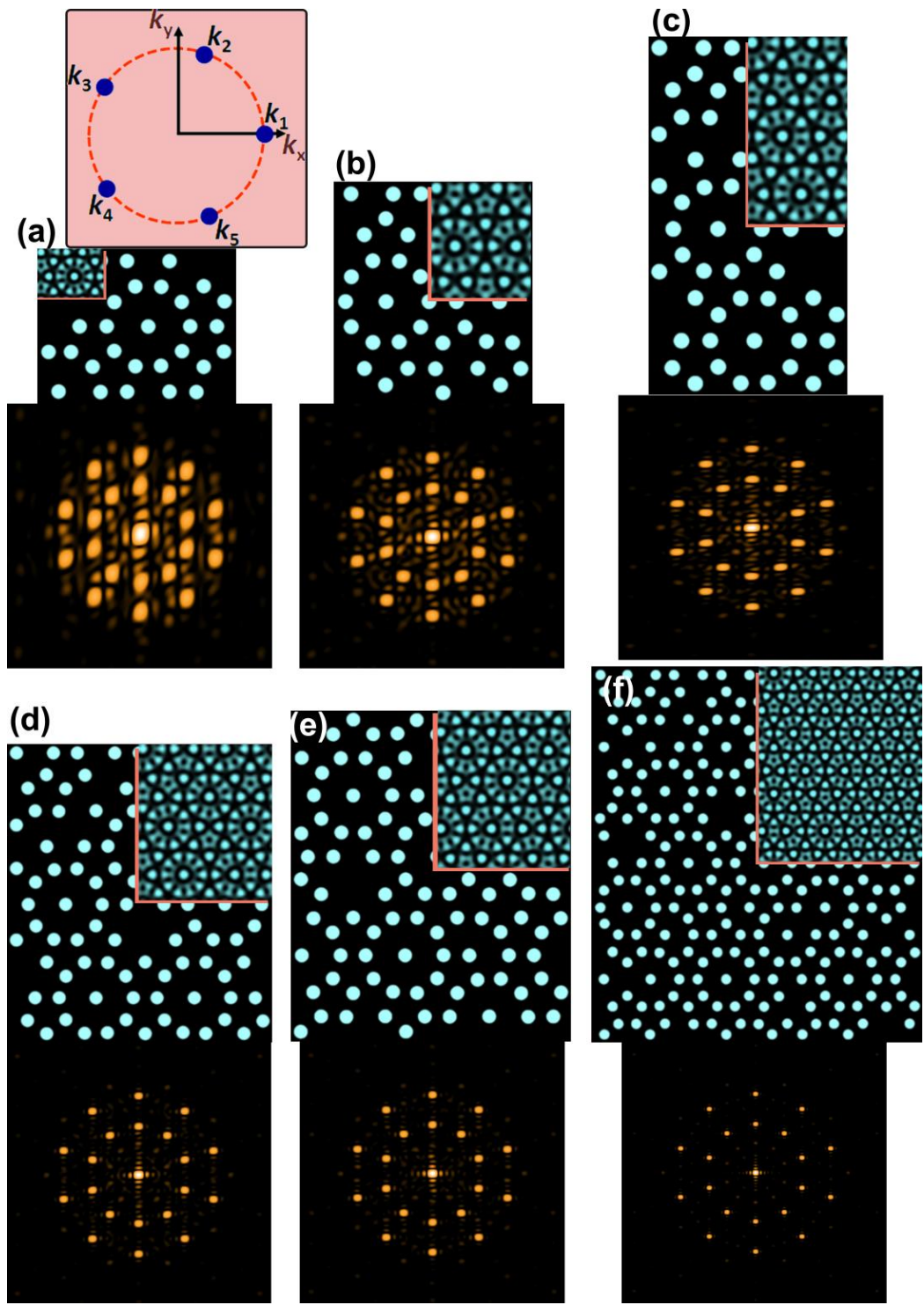

**Figure S2**

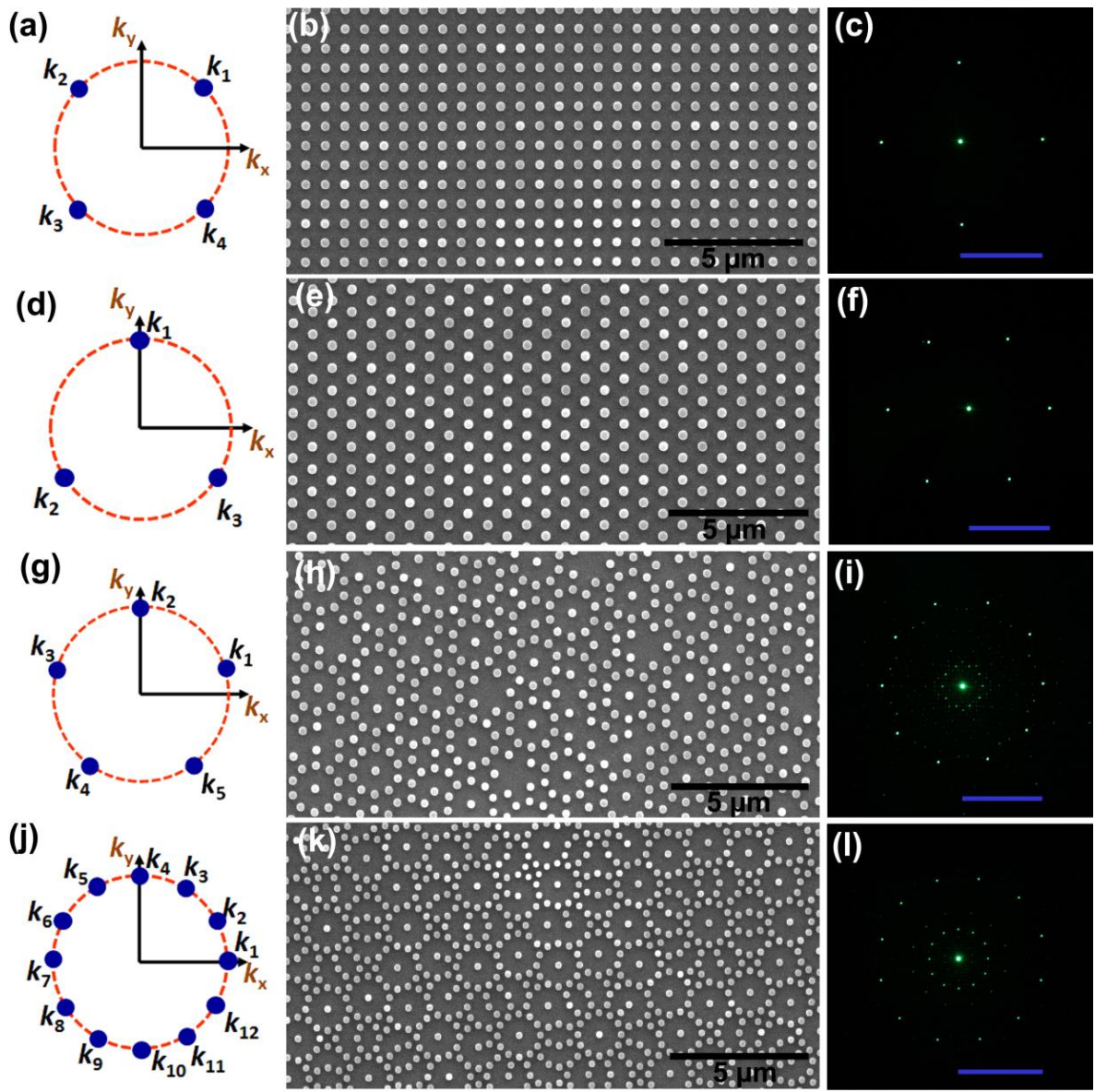

**Figure S3**

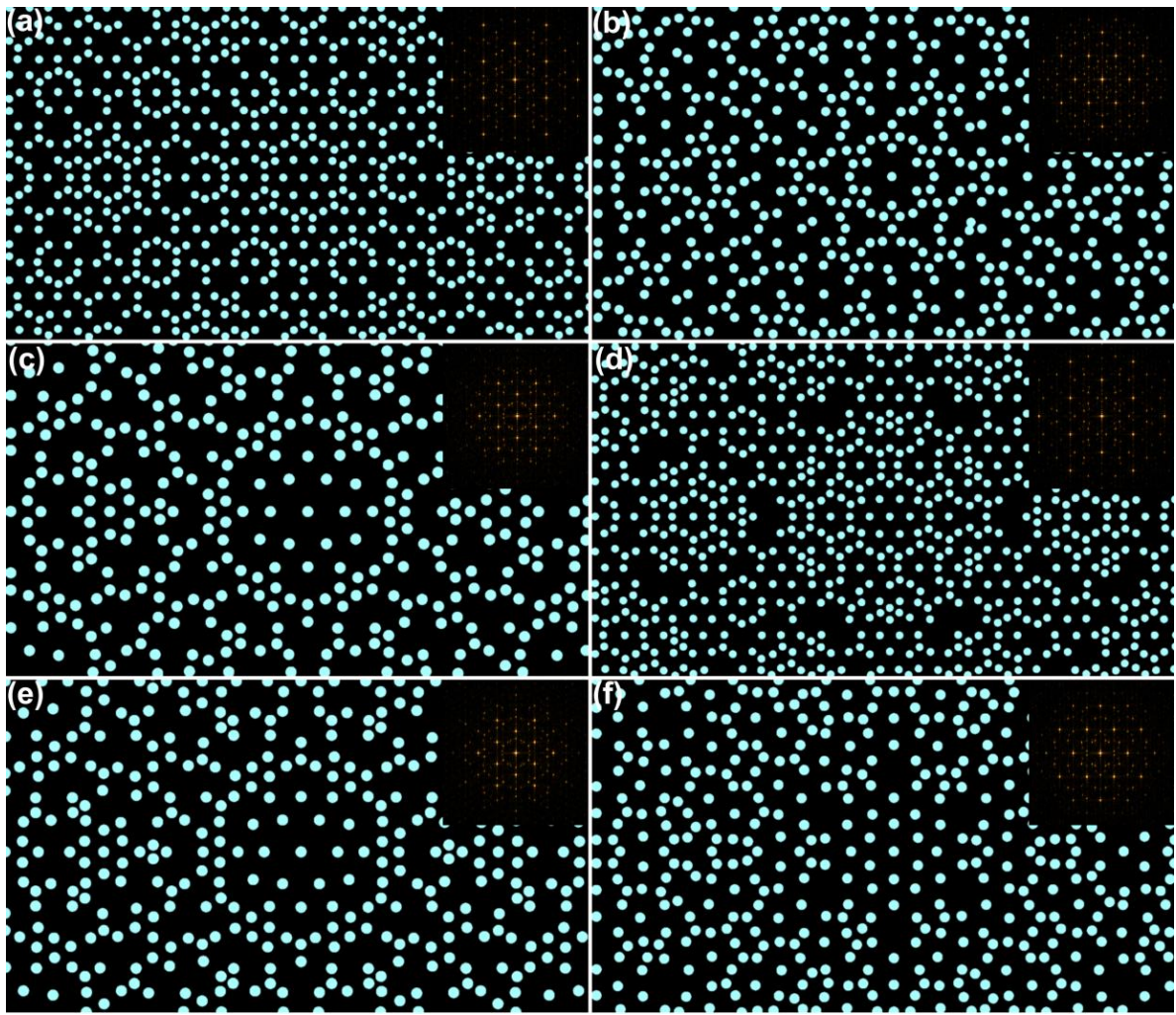

**Figure S4**

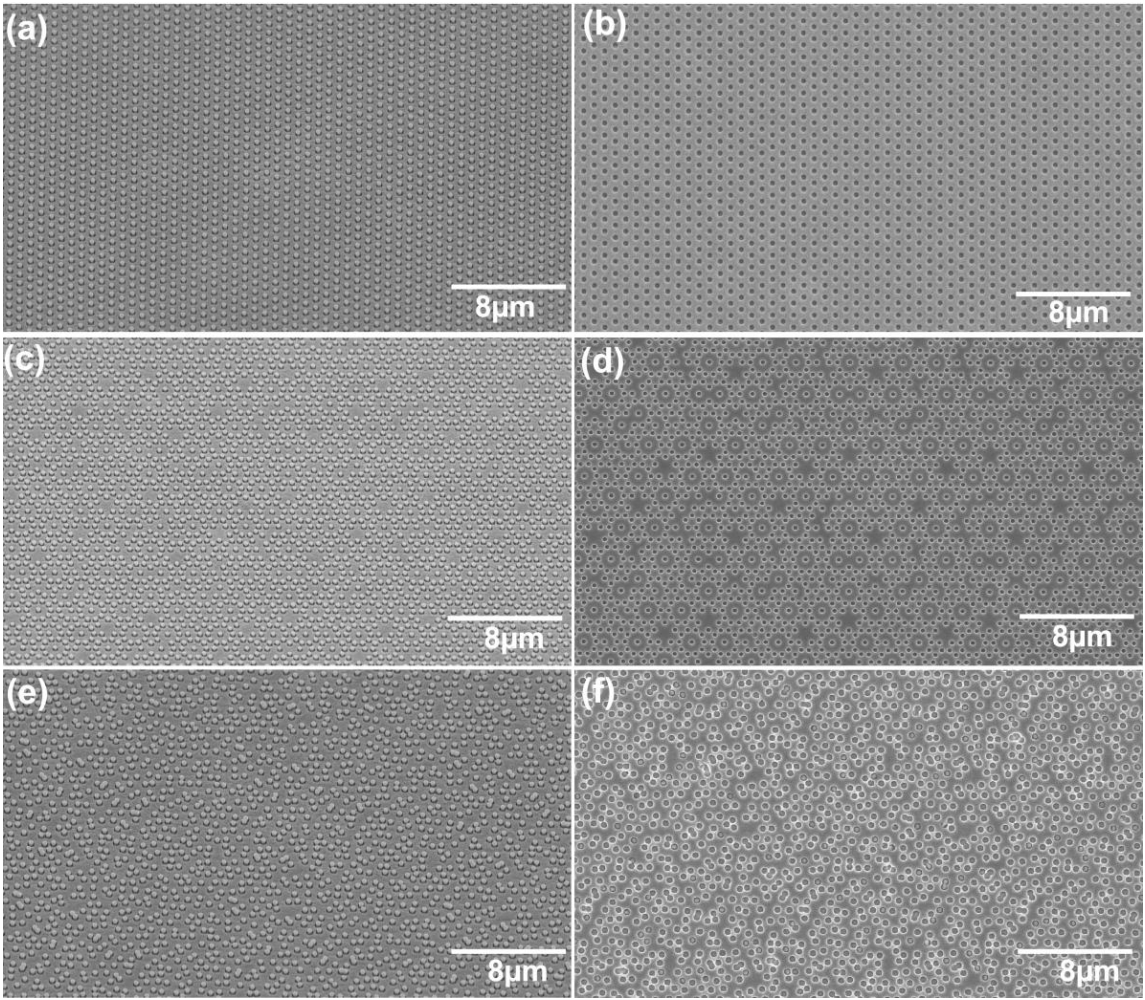